\documentstyle[11pt,epsfig,subeqn,cite]{article}
\def\be{\begin{equation}}
\def\bea{\begin{eqnarray}}
\def\ee{\end{equation}}
\def\eea{\end{eqnarray}}
                              \def\barr{\begin{array}}
                              \def\earr{\end{array}}
\def\dis{\displaystyle}

\def\etal{{\em et al.}}

                              \def\gev{\: \rm GeV}

\def\gappeq{\mathrel{\rlap {\raise.5ex\hbox{$>$}}
            {\lower.5ex\hbox{$\sim$}}}}
\def\lappeq{\mathrel{\rlap{\raise.5ex\hbox{$<$}}
            {\lower.5ex\hbox{$\sim$}}}}
\def\ra{\rightarrow}

\def\slashiii#1{\setbox0=\hbox{$#1$}#1\hskip-\wd0\hbox to\wd0{\hss\sl/\/\hss}}

\begin{document}                                                              
\begin{flushright}
MRI-P-001202\\
{\large \tt hep-ph/0012309}
\end{flushright}

\vspace*{2ex}

\begin{center}
{\Large\bf $B$-physics constraints on baryon number violating couplings: 
     grand unification or $R$-parity violation}

\vskip 15pt

{\sf Debrupa Chakraverty$^{1}$ and Debajyoti Choudhury${^2}$}

{\footnotesize Mehta Research Institute, 
Chhatnag Road, Jhusi, Allahabad 211 019, India. \\
E-mail: $^1$rupa@mri.ernet.in, $^2$debchou@mri.ernet.in}\\

\vskip 15pt
{\Large\bf Abstract}
\end{center}

We investigate the role that baryon number violating 
  interactions may play in $B$ phenomenology. 
 Present in various grand unified theories, supersymmetric theories with 
 $R$-parity violation and composite models,  a diquark state could be 
 quite light. Using 
 the data on $B$ decays as well as
 $B - {\bar B}$ mixing, we find strong constraints on the couplings that such a 
 light diquark state may have with the Standard Model quarks.  

\vskip 20pt


\vskip 20pt
\hspace*{0.65cm}

\def\baselinestretch{1.0}


\section{Introduction}
It is now widely accepted that the Standard Model (SM), despite its
great success, is only an effective theory.  The ills plaguing it may
be cured only in the context of a more fundamental theory operative at
higher energies.  The quest to find such a theory has, over the years,
inspired many a model going beyond the SM.  Two of the most attractive
classes of such models are those incorporating grand
unification~\cite{GUT} and/or supersymmetry~\cite{SUSY}.  The exact
nature of such a theory, however, is a matter of intense debate,
occasioned, not in the least, by the absence of any experimental
signature yet.  It is not surprising, thus, that the search for new
physics effects constitutes a major component of research in high
energy physics. Such efforts can be broadly classified into two
categories. On the one hand there are the direct searches typified by high
energy collider experiments where new particles are sought to be
produced on-shell and detected through their subsequent decays. The
other approach concentrates on indirect effects as can be deduced from
possible deviations from the SM predictions for low-energy
observables.  In this article we shall focus on one such set of low and
intermediate energy experiments.

The next decade will see a blossoming of experimental facilities
planning to explore $B -{\bar B}$ mixing  as well as $B$-meson decays 
with greater accuracy and for an increasing number of different final
states.  In light of these upcoming experiments (CLEO, BaBar, BELLE,
HERAB, BTEV and LHCB), it is of importance to examine their
sensitivity to new physics beyond the SM.

In this paper, we investigate the possible influence that a baryon
number violating interaction may have on $B$- phenomenology.  Within
the SM, baryon ($\hat B$) and lepton ($\hat L$) number conservations
come about due to accidental symmetries. In other words, such
conservations are not guaranteed by any principle, but are
rather the consequences of the choice of the particle
content~\footnote{Indeed, nonperturbative effects within the SM itself
do break $\hat B + \hat L$ symmetry.}.  In extensions of the SM, such
an accidental occurrence is obviously not guaranteed. For example,
even in the simplest grand unified theories (GUTs), both the gauge and
the scalar sector interactions violate each of $\hat B$ and $\hat
L$. The corresponding particles, namely the diquarks~\cite{diquarks}
and leptoquarks~\cite{leptoquarks} have been studied in the literature
to a considerable extent.

Simultaneous breaking of both $\hat B$ and $\hat L$ symmetry is
obviously a recipe for disaster as this combination is more than
likely to lead to rapid proton decay. Within GUTs, gauge
boson-mediated proton decay could be naturally suppressed by
postulating the symmetry breaking scale to be very large. However,
there do exist a class of GUTs~\cite{frampton}, where the next set of
thresholds need not be very high and $\hat B$-violating gauge
particles can be relatively light. Proton decay, however, remains
suppressed on account additional symmetries in the theory.
Suppression of the scalar mediated contribution to proton decay in a
generic GUT, on the other hand, is easier to obtain: the particle
content can be so chosen that there is no diquark-leptoquark mixing,
at least as far as the light sector is concerned.

In the case of the Minimal Supersymmetric Standard Model (MSSM),
though, we do not have the option of demanding the `offending' fields
(the supersymmetric partners of the SM fermions) to be
superheavy. Ruling out the undesirable terms necessitates the
introduction of a discrete symmetry, $R \equiv (-1)^{3 (\hat B - \hat
L) + 2 \hat S}$ (with $\hat S$ denoting the spin of the
field)~\cite{fayet}.  Apart from ruling out both $\hat B$ and $\hat L$
violating terms in the superpotential, this symmetry has the
additional consequence of rendering the lightest supersymmetric
partner absolutely stable. However, such a symmetry is {\em ad
hoc}. Hence, it is of interest to consider possible violations of this
symmetry especially since it has rather important experimental
consequences, not the least of which concerns the detection of the
supersymmetric partners.

It can thus be argued that, in such models as well as in models of
compositeness~\cite{diquarks}, it is quite likely that baryon number violating
interactions may not be suppressed too severely. Even more
interestingly, such processes may be mediated by relatively low-lying
states, generically called diquarks.

This paper is organised as follows: Section 2 
 constitutes  a brief review on diquarks. 
Section 3 deals with  hadronic $B$ decays. In section 4, we concentrate
 on  $B -{\bar B}$ mixing. Section 5 contains the numerical results.
 We conclude in section 6 with a summary and outlook.

\section{Diquarks: a brief review}

In this section we shall briefly examine all possible tree-level $\hat
B$-violating couplings involving the SM quarks. We shall adopt a
purely phenomenological standpoint without any particular reference or
prejudice to the origin of such couplings or states.  A generic
diquark is a scalar or vector particle that couples to a quark current
with a net baryon number ${\hat B} = \pm 2/ 3$. Clearly, under
$SU(3)_c$, it may transform as either a triplet or a sextet.  For
scalars, the Yukawa term in the Lagrangian can be expressed as
\subequations \be {\cal L}_{\rm SD} = h_{ij}^{(A)} {\bar q}_i^c
P_{L,R} q_j \Phi_A + h.c., \ee where $i,j$ denote quark flavours, $A$
denotes the diquark type and $P_{L,R}$ reflect the quark chirality.
Standard Model gauge invariance demands that a scalar diquark
transforms either as a triplet or as a singlet under $SU(2)_L$.  For a
vector diquark, on the other hand, the relevant term in the Lagrangian
can be parametrized as \be {\cal L}_{\rm VD} = \vartheta_{ij}^{(A)}
{\bar q}_i^c \gamma_\mu P_{L,R} q_j V_A^\mu + h.c.  \ee
\endsubequations with $V_A$ transforming as a $SU(2)_L$ doublet.  The
full list of quantum numbers, for either case, is presented in
Table~\ref{tab:qnos}.  Clearly, the couplings $h^{(1)}_{ij}$,
$h^{(4)}_{ij}$, $h^{(5)}_{ij}$ and $h^{(7)}_{ij}$ must be symmetric
under the exchange of $i$ and $j$ while $h^{(2)}_{ij}$,
$h^{(3)}_{ij}$, $h^{(6)}_{ij}$ and $h^{(8)}_{ij}$ must be
antisymmetric. For the other couplings, {\em viz.} $\tilde
h^{(3)}_{ij}$, $\tilde h^{(4)}_{ij}$, and $\vartheta^{(A)}_{ij}$, no
such symmetry property exists.  hereafter, we assume these couplings
to be real~\footnote{The extension to complex couplings is
straightforward. The imaginary parts, however, can be better
constrained from an analysis of the CP violating decay modes.}.  Note
that the quantum numbers of $\Phi_{2, 4, 6}$ as well as those of
$V^\mu_{2, 4}$ allow them to couple to a leptoquark ({\em i.e.} a
quark-lepton) current as well.  Clearly, the non-observance of proton
decay implies that such $L$-violating couplings must be suppressed
severely.

 It should be noted that we are not demanding that the vector 
 diquarks correspond to some gauge theory. While it might be 
 rightly argued that a theory with non-gauged
 vector particles is non-renormalizable, one should keep in mind that 
 such states may well be there in an effective theory. Since we would be
 studying the phenomenological implications only at the lowest order of
 perturbation theory, renormalizability is not an issue here.


\begin{table}[h]
\begin{center}
\begin{tabular}{|c|c|c|}
\hline
& &  \\[-2ex]
 Diquark Type  &  Coupling & $SU(3)_c\times SU(2)_L\times U(1)_Y$\\
& &  \\[-2ex]
 \hline
 \hline
& &  \\[-2ex]
 $\Phi_1$ & $h_{ij}^{(1)} ({\bar Q}_{L i})^c Q_{L j} \Phi_1$ & 
	($ {\bar 6}$, $3$, $-{2\over 3}$)\\ 
& &  \\[-2ex]
 \hline
& &  \\[-2ex]
 $\Phi_2$ & $h_{ij}^{(2)} ({\bar Q}_{L i})^c Q_{L j} \Phi_2$ & 
	($3$, $3$, $-{2\over 3}$)\\ 
& &  \\[-2ex]
\hline
& &  \\[-2ex]
 $\Phi_3$ & 
   $\left[ h_{ij}^{(3)} ({\bar Q}_{L i})^c Q_{L j} 
         + \tilde h_{ij}^{(3)}({\bar u}_{R i})^c d_{R j}
    \right] \: \Phi_3 $ & ($ {\bar 6}$, $1$, $-{2\over 3}$)\\ 
& &  \\[-2ex]
\hline
& &  \\[-2ex]
 $\Phi_4$ & 
    $\left[ h_{ij}^{(4)} ({\bar Q}_{L i})^c Q_{L j} 
          + \tilde h_{ij}^{(4)}({\bar u}_{R i})^c d_{R j}
     \right] \Phi_4 $ & ($ 3$, $1$, $-{2\over 3}$)\\
& &  \\[-2ex]
 \hline
& &  \\[-2ex]
 $\Phi_5$ & $h_{ij}^{(5)} ({\bar u}_{R i})^c u_{R j}
 \Phi_5$ & ($ {\bar 6}$, $1$, $-{8\over 3}$)\\
& &  \\[-2ex]
 \hline
& &  \\[-2ex]
 $\Phi_6$ & $h_{ij}^{(6)} ({\bar u}_{R i})^c u_{R j}
 \Phi_6$ & ($ 3$, $1$, $-{8\over 3}$)\\
& &  \\[-2ex]
 \hline
& &  \\[-2ex]
 $\Phi_7$ & $h_{ij}^{(7)} ({\bar d}_{R i})^c d_{R j}
 \Phi_7$ & ($ \bar 6$, $1$, ${4\over 3}$)\\
& &  \\[-2ex]
 \hline
& &  \\[-2ex]
 $\Phi_8$ & $h_{ij}^{(8)} ({\bar d}_{R i})^c d_{R j}
 \Phi_8$ & ($3$, $1$, ${4\over 3}$)
	\\
& &  \\[-2ex]
 \hline
& &  \\[-2ex]
 ${V_1}^\mu$ & $\vartheta_{ij}^{(1)} ({\bar Q}_{L i})^c \gamma_\mu d_{R j}
 V_1^\mu$ & ($\bar 6$, $2$, ${1\over 3}$)\\
& &  \\[-2ex]
 \hline
& &  \\[-2ex]
 ${V_2}^\mu$ & $\vartheta_{ij}^{(2)} ({\bar Q}_{L i})^c \gamma_\mu d_{R j}
 V_2^\mu$ & ($3$, $2$, ${1\over 3}$)\\
& &  \\[-2ex]
 \hline
& &  \\[-2ex]
 ${V_3}^\mu$ & $\vartheta_{ij}^{(3)} ({\bar Q}_{L i})^c \gamma_\mu u_{R j}
 V_3^\mu$ & ($\bar 6$, $2$, $-{5\over 3}$)\\
& &  \\[-2ex]
 \hline
& &  \\[-2ex]
 ${V_4}^\mu$ & $\vartheta_{ij}^{(4)} ({\bar Q}_{L i})^c \gamma_\mu u_{R j}
 V_4^\mu$ & ($3$, $2$, $-{5\over 3}$)\\
& &  \\
 \hline
\end{tabular}
\end{center}
    \caption{\em Gauge Quantum Numbers and Yukawa Couplings 
	         of Diquarks $(Q_{em} = T_3 +{Y \over 2})$.
            }
	\label{tab:qnos}
\end{table}


We now turn to the MSSM, where both $\hat B$-- and $\hat L$-violating
terms are allowed, in general, by supersymmetry as well as gauge
invariance. As stated earlier, catastrophic rates for proton decay can
be avoided by imposing a global $Z_2$ symmetry~\cite{fayet}
under which the quark and lepton superfields change by a sign, while
the Higgs superfields remain invariant. However, since such a symmetry
is entirely {\em ad hoc} within the purview of the MSSM, it is
conceivable that this $R$-parity may be broken while keeping either of
$\hat B$ or $\hat L$ intact.  In our study, we shall restrict
ourselves to the case where only the ${\hat B}$-violating terms are
non-zero. Such scenarios can be motivated from a class of
supersymmetric GUTs as well~\cite{Bviol}. The corresponding terms in
the superpotential can be parametrized as \be W_{R\!\!\!/} =
\lambda''_{ijk} \bar{U}^i_R \bar{D}^j_R \bar{D}^k_R, \label{superpot}
\ee where $\bar{U}^i_R$ and $\bar{D}^i_R$ denote the right-handed
up-quark and down-quark superfields respectively.  The couplings
$\lambda''_{ijk}$ are antisymmetric under the exchange of the last two
indices. The corresponding Lagrangian can then be written in terms of
the component fields as: \be {\cal L}_{R\!\!\!/} = \lambda''_{ijk}
\left(u^c_i d^c_j \tilde{d}^*_k + u^c_i \tilde{d}^*_j d^c_k +
\tilde{u}^*_i d^c_j d^c_k\right) + {\rm h.c.}
\label{lagrp}
\ee
Thus, a single term in the superpotential corresponds to  
two of type $\tilde h^{(4)}_{ij}$  and one of 
type $h^{(8)}_{ij}$ diquark interactions. 

The best direct bound on diquark type couplings is derived  from  an 
analysis of dijet events at the Tevatron~\cite{CDF_dijet}.
Considering the process $q_i q_j \ra \Phi_A \ra q_i q_j$, an exclusion
curve in the ($m_{\Phi_A}, h^{(A)}_{ij}$) plane can be obtained from
this data. A similar statement holds for the vector particles as
well. Two points need to be noted though.  At a $p \bar p$ collider
like the Tevatron, the $u u $ and $d d $ fluxes are small and
hence the bounds are relatively weak.  This is even more true for
quarks of the second or third generation (which are relevant for the
couplings that we are interested in). Secondly, such an analysis needs
to make assumptions regarding the branching fraction of $\Phi_A$
($V_A$) into quark pairs, a point that is of particular importance in
the context of $R$-parity violating supersymmetric models.

There also exist some constraints derived from low energy processes.
Third generation couplings, for example, can be constrained from the
precision electroweak data at LEP \cite{b_c_s} or, to an extent, by
demanding perturbative unitarity to a high scale~\cite{biswa}.
Couplings involving the first two generations, on the other hand, are
constrained~\footnote{Although many of these analyses have been done
for the case of $R$-parity violating models, clearly similar bounds
would also apply to nonsupersymmetric diquark couplings as well.}  by
the non-observance of neutron-antineutron oscillations or from an
analysis of rare nucleon and meson decays~\cite{sher,probir}.  While
many of these individual bounds are weak, certain of their {\em
products} are much more severely constrained by the data on neutral
meson mixing and $CP$--violation in the
$K$--sector~\cite{barbieri}. It is our aim, in this article, to derive
analogous but stronger bounds.

\begin{table}[h]
\begin{center}
\begin{tabular}{|c|c|}
\hline
&  \\[-2ex]
 Diquark Type  &  Effective Four Quark Operator \\
&  \\[-2ex]
 \hline
 \hline
&  \\[-2ex]
 $\Phi_1$ & 
     $\barr{rcl}
	\dis {{h_{ij}^{(1)}h_{kl}^{(1)}} \over {16 m_{\Phi_1}^2}} 
	& \Big[ & 
	\dis  ({\bar  u}_k \gamma_\mu L u_i) \: ({\bar d}_l \gamma^\mu L d_j) 
	    + ({\bar  d}_k \gamma_\mu L d_i) \: ({\bar u}_l \gamma^\mu L u_j) 
		\\
	& + & 
	\dis ({\bar  d}_k \gamma_\mu L u_i) \: ({\bar u}_l \gamma^\mu L d_j)
	+ ({\bar  u}_k \gamma_\mu L d_i) \: ({\bar d}_l \gamma^\mu L u_j)
	          \\
        &+& 
         2 \: ({\bar  d}_k \gamma_\mu L d_i) \: 
				({\bar d}_l \gamma^\mu L d_j) 
	\Big] + h.c.
	\earr
     $
\\
 \hline
&  \\[-2ex]
 &      $\barr{rcl}
	\dis {{h_{ij}^{(3)}h_{kl}^{(3)}} \over {8 m_{\Phi_3}^2}} 
	& \Big[ & 
	\dis ({\bar  u}_k \gamma_\mu L u_i) \: ({\bar d}_l \gamma^\mu L d_j)
	     + ({\bar  d}_k \gamma_\mu L d_i) \: ({\bar u}_l \gamma^\mu L u_j)
		\\
	& - & 
	\dis ({\bar  d}_k \gamma_\mu L u_i) \: ({\bar u}_l \gamma^\mu L d_j)
	     - ({\bar  u}_k \gamma_\mu L d_i) \: ({\bar d}_l \gamma^\mu L u_j)
	\Big] + h.c.
	\earr
     $
		\\
  \cline{2-2}
&  \\[-2ex]
$\Phi_3$ &  
     $
	\dis {{\tilde h_{ij}^{(3)} \tilde h_{kl}^{(3)}} 
		\over {8 m_{\Phi_3}^2}} \: 
	({\bar  u}_k \gamma_\mu R u_i) \: ({\bar d}_l \gamma^\mu R d_j)
      + h.c. $
		\\
 &  \\[-2ex]
  \cline{2-2}
&  \\[-2ex]
&    $\barr{rcl}  
      	\dis {{\tilde h_{ij}^{(3)} h_{kl}^{(3)}}
		\over {8 m_{\Phi_3}^2}}  
	   & \Big[ & 
  \dis ({\bar  d}_k R u_i) \: ({\bar u}_l R d_j)
	 -  ({\bar  u}_k R u_i) \: ({\bar d}_l R d_j) \\
   & + & {1 \over 4} \{ \dis  ({\bar  d}_k \sigma^{\mu \nu} R u_i) \: ({\bar u}_l \sigma_{\mu \nu} R d_j)
         -  ({\bar  u}_k \sigma^{\mu \nu} R u_i) \: ({\bar d}_l \sigma_{\mu \nu} R d_j)\} 
	     \Big] + h.c.
   \earr
     $
\\ 
 &  \\[-2ex]
 \hline
&  \\[-2ex]
$\Phi_7$ &  
$\dis {{h_{ij}^{(7)}h_{kl}^{(7)}} \over { 8 m_{\Phi_7}^2}}
  ({\bar  d}_k \gamma_\mu R d_i) \: ({\bar d}_l \gamma^\mu R d_j)  + h.c.$ \\
 &  \\[-2ex]
 \hline
&  \\[-2ex]
$V_1$ &   
 $ 
	\dis {{- \vartheta_{ij}^{(1)}\vartheta_{kl}^{(1)}} \over {4 m_{V_1}^2}}
		\:
	\Big[ ({\bar u}_k \gamma_\mu L u_i) +({\bar d}_k \gamma_\mu L d_i)
	\Big] \:  ({\bar d}_l \gamma^\mu R d_j) + h.c.
  $
\\ 
 &  \\[-2ex]
\hline
 &  \\[-2ex]
$V_3$ & 
 $\dis {{- \vartheta_{ij}^{(3)}\vartheta_{kl}^{(3)}} \over { 4 m_{V_3}^2}}
 ({\bar d}_k \gamma_\mu L d_i)({\bar u}_l \gamma^\mu R
u_j) + h.c. $\\[2ex]
\hline
\end{tabular}
\end{center}
    \caption{\em The effective four quark operator
	         for various diquarks. The operators for 
	         $\Phi_2$, $\Phi_4$, $\Phi_8$, $V_2$ and $V_4$ mirror
		 those for $\Phi_1$, $\Phi_3$, $\Phi_7$, $V_1$ and $V_3$
		 respectively, albeit with a different colour 
		 factor (see text). Here $L(R) =1 \mp \gamma_5$. 
            }
	\label{tab:vert}
\end{table}

At energy scales well below the mass of the diquark, the latter can be
integrated out and effective four quark operators obtained.  In
Table~\ref{tab:vert}, we list these for each diquark type.  A few
points should be noted here:
\begin{itemize}
	\item we have not displayed 
the operators resulting from $\Phi_5$ and $\Phi_6$ 
as these do not contribute (at the lowest order) to either 
 $B$ decays or $B - {\bar B}$ mixing;
	\item for convenience's sake, 
we have Fierz-rearranged the operators and, in the process, 
exchanged the charge-conjugated fermion fields (which come in 
naturally) for their non-conjugated counterparts; 
	\item within a diquark multiplet, we have assumed all the fields 
to be mass degenerate since large splittings within a multiplet are
anyway disfavoured by LEP data;
 \item We have neglected the evolution of the diquark mediated effective 
four-quark interactions from the electroweak
 scale down to $B$ meson scale through renormalisation group equations;  
	\item we have not displayed 
the extra color factors that appear on account of the 
diquark states being coloured objects. 
The said factors can be determined by reexpressing four
quark operators of the forms
$({\bar 3}_c \otimes 3_c)_1$ and $(6_c \otimes {\bar 6}_c)_1$ 
in terms of the corresponding 
$( 1_c \otimes 1_c)_1$ and $ (8_c \otimes 8_c)_1$ current structures. 
Thus, transforming 
$({\bar q}^c_i\Gamma q_j)  ({\bar q}_l \Gamma^\prime  q_k^c)$ 
to the form 
$({\bar q}_k \Gamma^{\prime \prime} q_i)
 ({\bar q}_l \Gamma^{\prime \prime \prime} q_j)$
implies that we are dealing with linear combinations of the form 
\be
\barr{rcl}
 ({\bar 3}_c \otimes 3_c)_1 & = & \dis {2\over 3}( 1_c \otimes 1_c)_1
 -( 8_c \otimes 8_c)_1 \\[2ex]
(6_c \otimes {\bar 6}_c)_1 & = & \dis {2\over 3}( 1_c \otimes 1_c)_1
 + {1 \over 2} (8_c \otimes 8_c)_1
\earr
	\label{color_struct}
\ee
These extra color-factors need to be included while calculating the 
hadronic matrix elements.  

\end{itemize}

Looking at Table~\ref{tab:vert}, it is obvious that the 
effective Hamiltonian for the full theory can be parametrized as 
\subequations
\be
    {\cal H}_{\rm eff} = \sum_{i=0}^9 b_i {\cal H}_i
\ee
with
\be
\barr{rclcrcl}
{\cal H}_0 & = & ({\bar q_1} \gamma_\mu R b) ({\bar q_2} \gamma^\mu R q_3)
	& \qquad & 
{\cal H}_1 & = & ({\bar q_1} \gamma_\mu L b) ({\bar q_2} \gamma^\mu L q_3)
	\\[1.5ex]
{\cal H}_2 & = & ({\bar q_1} \gamma_\mu R b) ({\bar q_2} \gamma^\mu L q_3)
	& \qquad & 
{\cal H}_3 & = & ({\bar q_1} \gamma_\mu L b) ({\bar q_2} \gamma^\mu R q_3)
	\\[1.5ex]
{\cal H}_4 & = & ({\bar q_1} L b) ({\bar q_2} R q_3)
	& \qquad & 
{\cal H}_5 & = & ({\bar q_1} R b) ({\bar q_2} L q_3) 
	\\[1.5ex]
{\cal H}_6 & = & ({\bar q_1} L b) ({\bar q_2}  L q_3)
	& \qquad & 
{\cal H}_7 & = & ({\bar q_1} R b) ({\bar q_2} R q_3)
	\\[1.5ex]
{\cal H}_8 & = & ({\bar q_1} \sigma_{\mu \nu} L b) ({\bar q_2} 
\sigma^{\mu \nu} L q_3)
	& \qquad & 
{\cal H}_9 & = & ({\bar q_1} \sigma_{\mu \nu} R b) ({\bar q_2} 
\sigma^{\mu \nu} R q_3).
\earr
	\label{effective}
\ee
\endsubequations
The strengths $b_i$ include both the SM contributions 
as well as diquark contributions (as in Table~\ref{tab:vert}) wherever 
applicable.

It now remains to calculate the hadronic matrix elements for ${\cal H}_i$,
a task that is rendered very difficult by the associated strong interaction 
dynamics. Hence, instead of attempting an exact calculation, one normally 
takes recourse to some appropriate approximation. In the 
``Naive Factorisation'' approach~\cite{ali}, the 
matrix elements of a four-quark operator are approximated by products
of matrix elements of the associated quark bilinears. As an  example, 
the amplitude for the decay $B \ra X_1 + X_2$ (where $X_{1,2}$ are arbitrary 
mesons) can be expressed as
\be
\barr{rcl}
\dis
\langle X_1 X_2 (\vert {\bar q_1}\Gamma b)
		({\bar q_2} \Gamma^\prime q_3)\vert B \rangle
&	\approx & \dis
     \langle X_1 \vert   {\bar q_1} \Gamma b  \vert B \rangle \:
           \langle X_2 \vert  {\bar q_2} \Gamma' q_3 \vert 0 \rangle 
     + 
     \langle X_2 \vert   {\bar q_1} \Gamma b  \vert B \rangle \:
           \langle X_1 \vert  {\bar q_2} \Gamma' q_3 \vert 0 \rangle 
  \\[1.5ex]
& + & 
     \frac{1}{N_c} \: 
	\langle X_1 \vert   {\bar q_1} \Gamma'' b  \vert B \rangle \:
           \langle X_2 \vert  {\bar q_2} \Gamma''' q_3 \vert 0 \rangle
 +  \frac{1}{N_c} \: 
	\langle X_2 \vert   {\bar q_1} \Gamma'' b  \vert B \rangle \:
           \langle X_1 \vert  {\bar q_2} \Gamma''' q_3 \vert 0 \rangle 
\earr     
	\label{factorization}
\ee
where the second line refers to Fierz rearranged currents.  Of course,
only some of the matrix elements are non-vanishing. For one, within
this approximation, the contributions of the tensor operators in
eq.(\ref{effective}) vanish identically.  It should be noted that
eqs.(\ref{effective} \& \ref{factorization}) 
contain only color-singlet currents.  For
color-octet currents to contribute, one would need to consider
additional gluon exchanges. Such effects are clearly not factorizable.
Within this approximation then, the color octet parts of
eq.(\ref{color_struct}) can be neglected, or, in other words, the
contribution of a color-sextet diquark is almost indistinguishable
from that of the corresponding color-triplet one.

\section{Hadronic $B$ decays}

Within the SM, hadronic $B$ decays may proceed through either tree
level $W$ boson exchange diagrams and/or through penguin diagrams
(both QCD and electroweak). The corresponding effective Hamiltonian,
including the QCD corrections have been presented in
Refs.\cite{ali,gatto}.  For brevity's sake, we do not repeat the
entire list here.  It suffices to remember that the SM amplitudes are
proportional to the Fermi constant $G_F$, the relevant product of two
CKM matrix elements $V_{ib}V^*_{jk}$ and/or $V_{tb}V^*_{tk}$ (with $i$
and $j$ as generic up type quarks and $k$ as down type quark) and the
combination of the Wilson coefficients that incorporates the short
distance QCD corrections at the $B$ mass scale.  The considerable
suppression due to the smallness of the CKM mixing is what makes
$B$-decays sensitive to new physics effects.

Reverting to the calculation of the hadronic matrix elements, 
the decay constant $f_i$ for a generic (pseudoscalar or vector) 
meson is defined through the relations
\be
\barr{rcl}
\langle P(p_P) \vert {\bar q}_j \gamma_\mu \gamma_5 q_i \vert 0 \rangle 
  & = & -if_P p_P^\mu\\
\langle V(p_V) \vert {\bar q}_j \gamma_\mu  q_i \vert 0 \rangle 
 & = & f_V m_V \epsilon^\mu.
\earr
	\label{decay_const}
\ee
Here it is assumed that the meson is composed of a $q_j \bar q_i$ pair.
The decay constants are best determined from an analysis 
of the respective leptonic decay modes and the relevant ones are listed in 
Table~\ref{tab:decay}.
The matrix elements for the associated density operators may then be 
evaluated using the Dirac equation:
\[
\barr{rcl}
\partial^\alpha({\bar q_i} \gamma_\alpha q_j) & = & i(m_j - m_i) {\bar q_i}q_j
\\
\partial^\alpha({\bar q_i} \gamma_\alpha\gamma_5  q_j) & = &
i(m_j +m_i) {\bar q_i} \gamma_5 q_j.
\earr
\]

\begin{table}[h]
\begin{center}
\begin{tabular}{|c|c|c|c|c|c|c|c|c|}
\hline
& & & & & & & & \\[-2ex]
$ f_\pi$ & $f_\rho$ & $f_K$ & $f_{K^*}$ & $f_D$ & 
 $f_{D^*}$ & $f_{D_s}$ & $f_{D_s^*}$ &$f_{J/\Psi}$ \\
 & & & & & & & & \\[-2ex]
 \hline
$131$ &  $207$ 
& $158$ & $214$
& $200$ & $230$ & $250$ & $275$ & $405$  \\
 & & & & & & & &  \\
\hline
\end{tabular}
\end{center}
    \caption{\em Values of Decay Constants in MeV.
            }
	\label{tab:decay}
\end{table}


The matrix elements for quark bilinears between a
$B$ meson and a pseudoscalar/vector meson can be parametrized in 
terms of form factors:
\be
\barr{rcl}
\langle P(p_P) \vert {\bar q}_j \gamma_\mu (1 - \gamma_5) b \vert B(p_B)
 \rangle 
& = & \dis
\left [ (p_B + p_P)_\mu - {{m_B^2 - m_P^2}  \over q^2} q_\mu
 \right ] F_1(q^2)
 +  {{m_B^2 - m_P^2}  \over q^2} q_\mu F_0(q^2)
\\[2.5ex]
\langle V(p_V) \vert {\bar q}_j \gamma_\mu (1 - \gamma_5) b 
	\vert B(p_B) \rangle 
& = & \dis 
-\epsilon_{\mu \nu \alpha \beta} \epsilon^{\nu *}
 p_B^\alpha p_V^\beta {{2 V(q^2)}\over {(m_B + m_V)}}
-i(\epsilon^*_\mu - {{\epsilon^* \cdot q} \over q^2} q_\mu)
 ( m_B + m_V) A_1(q^2)
	\\[2.0ex]
&+ & \dis i \left( (p_B + p_V)_\mu - {{(m_B^2 - M_V^2)} \over q^2} q_\mu\right)
  {{(\epsilon^* \cdot q) \: A_2(q^2)} \over {(m_B + m_V)}} 
         \\[2.0ex]
&-& i {{2 m_V (\epsilon^*
 \cdot q)} \over q^2} q_\mu A_0(q^2),
\earr
	\label{form_fac}
\ee
where $q=p_B -p_{P(V)}$ and $\epsilon$ is the polarisation vector of $V$.
The apparent poles at $q^2 =0$ are fictitious since
\[
\barr{rcl}
F_1(0) & = & F_0(0)\\
2 m_V A_0(0) & = & (m_B + m_V) A_1(0) - (m_B-m_V)A_2(0) \ .
\earr
\]
The numerical values of the form factors can be calculated within a 
given model. For our analysis, we adopt the 
BSW model~\cite{bsw,neubert}, and the relevant form factors, at zero momentum 
transfer, are given in Table~\ref{tab:bsw}~\cite{ali,bsw}.
 It can easily be checked that choosing a different model for the calculation
 of hadronic matrix elements would not change our results appreciably.
\begin{table}[h]
\begin{center}
\begin{tabular}{|c|c|c|c|c|c|c|}
\hline
& & & & & & \\[-2ex]
 Decay Mode  &  $F_1(0)$ & $F_0(0)$ & $V(0)$ & $A_1(0)$ & $A_2(0)$ & $A_0(0)$\\
& & & & & & \\[-2ex]
 \hline
& & & & & & \\[-2ex]
$B \to \pi$ & $0.33$ & $0.33$ & & & &\\
& & & & & & \\[-2ex]
 \hline
& & & & & & \\[-2ex]
$B \to K$ & $0.38$ & $0.38$ & & & &\\
& & & & & &  \\[-2ex]
\hline
& & & & & & \\[-2ex]
 $ B \to D$ & $0.69$ & $0.69$ & & & &\\
& & & & & &  \\[-2ex]
\hline
& & & & & & \\[-2ex]
$B \to \rho$ & & & $0.33$ & $0.28$ & $0.28$ & $0.28$\\
& & & & & &  \\[-2ex]
\hline
& & & & & & \\[-2ex]
$B \to K^*$ & & & $0.37$ & $0.33$ & $0.33$ & $0.32$\\
& & & & & &  \\[-2ex]
\hline
& & & & & & \\[-2ex]
$B \to D^*$ & & & $0.71$ & $0.65$ & $0.69$ & $0.62$\\
& & & & & &  \\
\hline
\end{tabular}
\end{center}
    \caption{\em Form Factors at Zero Momentum Transfer in the BSW Model  }
	\label{tab:bsw}
\end{table}

For the $q^2$ dependence of these form factors we assume a simple 
pole formula~\cite{ali,bsw}
$F(q^2) = F(0)/(1 - q^2/m_{pole}^2)$ with the 
pole mass $m_{pole}$ the same as that of the 
lowest lying meson with the appropriate quantum numbers  ($J^P=0^+$ for
$F_0$; $1^-$ for $F_1$ and $V$; $1^+$ for $A_1$ and $A_2$; $0^-$ for $A_0$).
The values of these pole masses are presented in Table~\ref{tab:pole}~\cite{ali,bsw}.

\begin{table}[h]
\begin{center}
\begin{tabular}{|c|c|c|c|c|}
\hline
 & & & &\\[-2ex]
Current & $m(0^-)$ & $m(1^-)$ & $m(1^+)$ & $m(0^+)$\\
  & & & &\\[-2ex]
 \hline
 & & & &\\[-2ex]
 ${\bar u}b$ & $5.28$ & $5.32$ & $5.37$ & $5.73$\\
  & & & &\\[-2ex]
 \hline
 & & & &\\[-2ex]
 ${\bar d}b$ & $5.28$ & $5.32$ & $5.37$ & $5.73$\\
 & & & &\\[-2ex]
 \hline
 & & & &\\[-2ex]
 ${\bar s}b$ & $5.37$ & $5.41$ & $5.82$ & $5.89$\\
& & & &\\[-2ex]
 \hline
 & & & &\\[-2ex]
${\bar c}b$ & $6.30$ & $6.34$ & $6.73$ & $6.80$\\
& & & &\\
 \hline
\end{tabular}
\end{center}
    \caption{\em Values of Pole Masses in GeV.}
	\label{tab:pole}
\end{table}


With eqns.(\ref{decay_const} \& \ref{form_fac}) in place, 
calculation of the full matrix elements, within the factorisation 
approximation, is now a straightforward task. 
Consider the decay
$ B(b \bar q_4) \to M_1(q_1 \bar q_4) M_2(q_2 \bar q_3)$ where 
$M_i$ are generic mesons (pseudoscalar or vector). For simplicity's 
sake, assume that no two quarks are identical so that the quark bilinears
(see eq.\ref{factorization}) cannot relate the $B$ to $M_2$. In this case, 
the amplitudes are given by
\be
\barr{rcl}
\dis 
{\cal A} \left[B \to P_1 P_2 \right]
   & = & \dis 
	i f_{P_2} (m_B^2 -m_{P_1}^2) \; F_0^{B \to P_1}(m_{P_2}^2) \;
 \left[-b_0+b_1+b_2-b_3
  -{{(b_4 -b_5-b_6+b_7)m_{P_2}^2} \over {(m_b -m_{q_1})(m_{q_2} 
	+ m_{q_3})}}\right]
	\\[3ex]
\dis {\cal A} \left[B \to P_1 V_2\right] 
 & = & \dis 
2 f_{V_2} m_{V_2}  \; F_1^{B \to P}(m_{V_2}^2) \; 
 [b_0+b_1+b_2+b_3] \; (\epsilon^* \cdot p_{P_1})
	\\[2ex]
{\cal A} \left[B \to V_1 P_2\right] 
 & = & \dis 
  2 m_{V_1}  f_{P_2} \; (\epsilon^* \cdot p_{P_2}) \; 
	A_0^{B\to V_1}(m_{P_2}^2) \;
 \left[ b_0+b_1-b_2-b_3 
 +{{(b_4 +b_5-b_6 -b_7)m_{P_2}^2} \over {(m_b +m_{q_1})(m_{q_2} +
 m_{q_3})}} \right]
	\\[3ex]
{\cal A} \left[B \to V_1 V_2 \right] & = & \dis 
	f_{V_2} m_{V_2} \Bigg[
		 -\epsilon_{\mu \nu \alpha \beta} 
		\epsilon^{\mu *}_2 \epsilon^\nu_1 p_B^\alpha p_{V_1}^\beta 
			{{V(m_{V_2}^2)} \over {(m_B +m_{V_1})}} 
			(b_0+b_1+b_2+b_3)  \\[1ex]
&& \dis \hspace*{4em}
 -i (\epsilon^*_1 \cdot \epsilon_2) (m_B+m_{V_1}) A_1(m_{V_2}^2) \: 
		(-b_0+b_1 -b_2+b_3) \\
& & \dis \hspace*{4em}
	+ 2 i (p_B \cdot \epsilon_2) (p_B \cdot \epsilon^*_1)
 {{A_2(m_{V_2}^2)} \over {( m_B+ m_{V_1})}} \: (-b_0 + b_1-b_2 +b_3)
	\Bigg]
\earr
	\label{amplitudes}
\ee
For decay modes wherein $q_3 = q_4$, the second set amplitudes 
in eq.(\ref{factorization}) contribute too. These additional 
pieces, however, can 
be easily read off from eq.(\ref{amplitudes}). 

\section {$B^0 -{\bar B}^0$ Mixing}

 The main motivation for considering $B^0_d - {\bar B}^0_d$ mixing 
to constrain
 diquark couplings is that this mixing is mediated by 
flavour changing neutral current, which is forbidden  at the tree level in SM. 
The mixing is characterised by the experimentally 
 measurable mass difference 
\be
\Delta M_d = m_{B_d}^H - m_{B_d}^L  = {{\vert \langle {\bar B}^0_d \vert
{\cal H}_{\rm eff} \vert B^0_d \rangle \vert} \over {m_{B^0_d}}} 
\ee
with H and L denoting heavy and light mass eigen states. 
 The recent world average value of $\Delta M_d$ at $1 \sigma$ limit 
is~\cite{pdg}:
\be
\Delta M_d = (0.472\pm0.017)\times 10^{-12}s^{-1}.  
\ee

In SM, this mixing proceeds through the box diagrams with internal 
top quark and $W$ boson exchanges ~\cite{bbbar,buras}. 
The diagrams in which 
one or both top quarks are replaced by up or charm quarks are
negligible   
 on account of: 
($i$) the small mixing angles and ($ii$) the corresponding loop 
integrals being suppressed to a great extent due to the smallness of the 
light quark masses. 
 Integrating out the internal particles, one thus gets an
 effective four quark interaction,  with a $(V-A) \otimes (V-A)$ current
 structure, and scaling as 
 to $m_W^2 G_F^2 \vert V_{tb}V^*_{tk} \vert^2$. 
 The short distance QCD corrections are well determined~\cite{bbbar,buras},
 while the long-distance corrections are estimated to small, unlike in the case  of $K-{\bar K}$ mixing. 

 In presence of diquarks, two different types of contributions may appear. If there exist $\Delta b=\Delta d =2$ operators, then such a mixing can occur at the tree level itself.
 Else, additional contributions may appear in the form of new diquark-mediated
 box diagrams. However, as we are interested in small diquark couplings,
 we shall confine ourselves to tee level (in diquarks) processes only.

 In the  calculation of  the hadronic matrix element $\langle 
{\bar B}^0_d \vert {\cal H}_{i}
 \vert B^0_d \rangle$,
 the vacuum saturation approximation is a convenient one. Herein,  
 one inserts a complete set of states between the two currents and 
 assumes that the sum is dominated by the vacuum and thus the hadronic 
 matrix elements are proportional to $f_B^2$ by 
virtue of eq.(\ref{decay_const}). 
 The bag factor, $B_B$, 
 introduced to parametrize all possible deviations from the vacuum saturation
 approximation, can be  evaluated in various nonperturbative 
 approaches. We use here  the values of $B_B(\mu_b)$  
 and $f_B$ as obtained by UKQCD
 collaboration in  a quenched lattice calculation ~\cite{ukqcd}.
   
 Incorporating the contributions of all such current structures, 
  in addition to that of the SM, 
  we obtain
\be
\Delta M_d = f_B^2 B_B m_{B_d^0} \left\vert b_0+b_1 -b_2 -b_3 + {m_{B_d^0}^2
 \over {(m_b +m_d)^2}} (b_4 + b_5-b_6-b_7) \right\vert
\ee

This may then be compared with the experimental value to obtain 
the required constraints.

\section{Results}

Before we determine the bounds obtainable from $B$-phenomenology, 
it is worthwhile to reexamine the SM predictions for the decay modes
of interest; this helps in selecting the channels 
likely to result in stronger constraints. 
As mentioned earlier, within the SM, the hadronic
$B$ decays are mediated by one or more of tree ($W$-mediated), 
electroweak penguin and QCD penguin
diagrams. The branching fractions 
are determined primarily by the CKM mixings 
operative in the particular decay and, in case of one-loop processes, by the 
corresponding loop integral. For example, the decays 
$B^- \to D_s^- \pi^0$ and ${\bar B_d}^0 \to \pi^- \pi^+$ are suppressed 
in comparison to ${\bar B_d}^0 \to D^+ \pi^-, D^+ D_s^-$,  
	     $B^- \to D^0 D_s^-$ and  
	     ${\bar B_s}^0 \to D^+_s \pi^-$
on account of $V_{u b}$ being much smaller than $V_{c b}$. 
Similar statements obviously hold for the decays 
into the corresponding excited states. Inspite of such a suppression,
 the tree 
diagram far outweighs the one-loop contributions for any of these decays.
For the decays $B^- \to K^- \pi^0$, ${\bar B}^0
\to {\bar K}^0 \pi^0$ (and the corresponding $PV$ and $VV$ modes) though,
the tree level contributions 
are double Cabibbo suppressed with the consequence that 
these decays are dominated by penguin 
diagrams~\footnote{In our numerical calculations, we have used the 
	 Wilson coefficients as listed in Ref.~\protect\cite{ali} 
	for $N_c=3$.}. 
Decays like $B^- \to K^-K^0$, $B^- \to \pi^- {\bar K}^0$, on the 
other hand, are governed solely 
by electroweak and/or QCD penguins. 

A different suppression occurs for the processes ${\bar B}^0 \to \pi^0
\pi^0$, ${\bar B^0} \to D^0 \pi^0$.  Compared to the analogous modes
${\bar B}^0 \to \pi^+ \pi^-$, ${\bar B_d^0} \to D^+ \pi^-$ wherein the
charged mesons are created from the vacuum by a color-singlet current,
these decays are obviously color-suppressed.  In fact, the respective
short distance coefficients differ by as much as a factor of
$20$~\cite{ali}. 
And finally, there are the annihilation diagrams
  in the decays like
 $B(b {\bar q}) \to X_1(q_1 {\bar q}) X_2(q_2 {\bar q_1})$. 
  In these decays,
 $b$ and ${\bar q}$ in $B$ meson annihilate to produce ${\bar q}$  and $q_2$
 quarks, which, in turn form final state mesons with 
  $q_1 {\bar q}_1$ pair, created from vacuum.
 As these contributions are proportional to
the wavefunction at zero, 
they are typically much smaller than either
of the tree or penguin mediated spectator contributions.  For example,
Ref.~\cite{ali} argues that the annihilation amplitude for $B \to P_1
P_2$ modes is proportional to the mass difference of the mesons in the
final state and hence there is essentially no annihilation
contribution to ${\bar B} \to \pi^0 \pi^0$, ${\bar B}^0 \to K^+ K^-$
etc.

It is tempting to assume that the modes suppressed within the SM would
be the ones most sensitive to effects from new physics. While this is
largely so, a few points should be remembered. For one, color
suppression and/or suppression of annihilation diagrams are
essentially independent of weak matrix elements, and hence equally
applicable to either the SM or a theory with diquarks. Secondly, even
for decays wherein the SM amplitudes are Cabibbo suppressed, the
experimental data may not be precise enough for it to be a very
sensitive probe.  Rather, it could well turn out that an unsuppressed
mode may turn out to be one of the most sensitive on account of the
observations matching very well with the SM predictions.

In obtaining numerical results, we assume that 
only one pair of diquark couplings are nonzero. While this 
restriction may seem unwarranted, it is an useful approximation 
that allows one a quantitative appreciation of the various 
experimental constraints. Furthermore, we assume a 
common mass of $100 \gev$ for all the diquarks. As explained earlier, 
mass splittings between states in a single multiplet is disfavoured by 
LEP data. And since the effective four-Fermi operator goes 
as $m_{\Phi(V)}^{-2}$, for a general diquark mass, 
all our bounds on the products need only be rescaled by a 
factor of $( m_{\Phi(V)} / 100 \gev)^2$. 

The bounds, as obtained from a given decay mode, can be broadly 
classified into two sets. An experimentally observed channel
(with an associated error bar) would, in general, allow the 
 diquark coupling pairs to lie in one of two non-contiguous 
windows, with the separation between the windows determined by 
the agreement of the SM contribution. On the other hand, decay 
modes that are yet to be experimentally seen, can only lead 
to a single window. For a specific combination of diquark couplings, 
we look at all such individual bounds and then delineate the 
range satisfied by each. As an illustration, let us consider the product 
$h_{13}^{(1)} h_{12}^{(1)}$. At 90\% C.L., the ranges 
allowed by individual decays are as follows:
\[
\barr{llclcl}
(a): \qquad & {\bar B}^0 & \to & D^+ \rho^- & \quad : \quad & 
	[-6.0 \times 10^{-2}, -5.3 \times 10^{-2}], \ 
	[-7.3 \times 10^{-3},  5.7 \times 10^{-4} ]
	\\[1ex]
(b): \qquad & B^- & \to & D^0 \rho^-  & \quad : \quad & 
	[-4.7\times 10^{-2}, -4.2 \times 10^{-2}], \ 
	[-4.7 \times 10^{-4}, 4.8 \times 10^{-3}]
	\\[1ex]	
(c): \qquad & B^- & \to & {\bar K}^{0*} \pi^- & \quad : \quad & 
	[-8.9 \times 10^{-4}, 2.2 \times 10^{-3}]
	\\[1ex]
(d): \qquad & {\bar B}^0 & \to & {\bar K}^0 \rho^0 & \quad : \quad & 
	[-2.1 \times 10^{-3}, 1.6 \times 10^{-3}]
\earr
\]
Clearly the first window allowed by $(a)$ is completely ruled out by
the the bounds from $(c)$ and $(d)$. The same is true for the first
window from $(b)$.  Progressively eliminating parts of the domains
allowed by the individual decays, we find that the actual allowed
range for this particular combination is only $[-4.7 \times 10^{-4},
5.7 \times 10^{-4}]$. An identical strategy is adopted 
for all other combinations, and we list the best bounds
in Tables ~\ref{tab:p1}--\ref{tab:v3}. 

A very important point is to be noted here. In the preceding analysis,
while we have selected the range of parameters common to each
constraint, we have not really used the entire information available
to us. Such an analysis would involve the use of a statistical
discriminator such as a $\chi^2$ test or a likelihood test. While such
an exercise is a straightforward one and would have led to bounds
stricter than those we list, the decision to forego it was a conscious
one. For, in the absence of higher order corrections and a more
precise calculation of the hadronic matrix elements, the bounds 
derived here are only indicative. Hence, further refinement using
statistical methods is not really called for.


 From Table~\ref{tab:vert}, it is easy to see 
 that $ h_{ij}^{(1)}h_{kl}^{(1)}$, $h_{ij}^{(2)}h_{kl}^{(2)}$,
 $h_{ij}^{(3)}h_{kl}^{(3)}$,  $h_{ij}^{(4)}h_{kl}^{(4)}$, 
$h_{ij}^{(7)}h_{kl}^{(7)}$ and
$h_{ij}^{(8)}h_{kl}^{(8)}$ result in  
both neutral and charged current structures.
In a given hadronic decay, both the operators contribute with one of
 them being color suppressed. 
Furthermore, the flavor structure determines whether two contributions interfere
 constructively or destructively. 
  As a consequence, 
  the bounds on $h_{ij}^{(2)}h_{kl}^{(2)}$, $h_{ij}^{(4)}h_{kl}^{(4)}$
and $h_{ij}^{(8)}h_{kl}^{(8)}$  are weaker  by a factor of
 ${N_c + 1} \over {N_c - 1}$ compared to those for 
$h_{ij}^{(1)}h_{kl}^{(1)}$, $h_{ij}^{(3)}h_{kl}^{(3)}$ and
$h_{ij}^{(7)}h_{kl}^{(7)}$.
 A color-unsuppressed operator associated with combinations 
${\tilde h}^{(3)}_{ij}{\tilde h}^{(3)}_{kl}$ and
${\tilde h}^{(4)}_{ij}{\tilde h}^{(4)}_{kl}$ 
 are neutral current ones, 
 their contributions to charged current decays are naturally
 color-suppressed. Hence the corresponding bounds are weaker.
 Similarly, since ${h}^{(3)}_{ij}{\tilde h}^{(3)}_{kl}$ and
${h}^{(4)}_{ij}{\tilde h}^{(4)}_{kl}$,  are associated only with
 scalar, pseudoscalar and tensor operators, they cannot contribute
 to $B \to VV$ decays. And finally, as the diquarks $\Phi_{7,8}$
 couple only to down-type quarks, they can contribute only to 
 those decays that occur in the SM solely through penguin diagrams.
 
 The diquark $\Phi_3$ differs from $\Phi_4$ only by colour
quantum number.  But since the color factors for triplet and sextet
diquarks are accidentally equal and ${\tilde h}_{ij}^{(3)}$ and
${\tilde h}_{ij}^{(4)}$'s have no specific symmetry property under
the exchange of $i$ and $j$, the bounds on the product of these
couplings are exactly the same.  A similar story obtains for other
sets of diquarks, $(V_1, V_2)$ and $(V_3, V_4)$.

As discussed earlier, ${\tilde h}^{(4)}_{ij}$ is analogous to the
trilinear R parity violating coupling ${\lambda}^{\prime
\prime}_{ijk}$. Thus the constraints on ${\tilde h}^{(4)}_{ij}{\tilde
h}^{(4)}_{kl}$ are equivalent to those on 
${\lambda}^{\prime \prime}_{imj} {\lambda}^{\prime
\prime}_{kml}$. The upper bound on ${\tilde h}^{(4)}_{23}{\tilde
h}^{(4)}_{22}$ is marginally weaker than that, quoted
in~\cite{probir} and~\cite{ggw}. The upper bound on ${\tilde
h}^{(4)}_{11}{\tilde h}^{(4)}_{13}$ is much weaker than that listed
in~\cite{probir} and~\cite{ggw} . For other ${\tilde
h}^{(4)}_{ij}{\tilde h}^{(4)}_{kl}$, we obtain more stringent bounds.
For most of the nonsupersymmetric product of two diquark
couplings, our predicted bounds are much stronger.

In a few hadronic decays ($B^- \to K^- J/\Psi$, $B^- \to \pi^-
J/\Psi$ etc.), the SM predictions  do not agree with the experimental
 observation even at $2 \sigma$ level. While this could be construed 
 as an indication of new physics (in our case diquarks), we prefer to
 tread a more conservative path. 
Consequently, we have not included such decays in our analysis.

Turning to $B_d^0$--${\bar B}^0_d$ mixing, it is obvious that 
a tree-level contribution will accrue from any four-quark operator 
that violates both $b$-- and $d$--number by two units each. 
 Furthermore,
$\Phi_2$, $\Phi_3$ and $\Phi_8$ do not contribute to $B_d^0-{\bar
B}_d^0$ mixing by virtue of the antisymmetric nature of their
couplings under exchange of the flavor indices.  Thus $B_d^0-{\bar
B}_d^0$ mixing only imposes limits on the parameter space of
$\Phi_1$, $\Phi_7$, $V_1$ and $V_2$ diquarks. Unlike in the SM, $B_d^0-{\bar
B}_d^0$ occurs at tree level in presence of nonzero diquark
couplings. Accordingly, we obtain most stringent bounds on
 $h_{33}^{(A)}h_{11}^{(A)}$ and
$\vartheta_{33}^{(A)} \vartheta_{11}^{(A)}$.

\section{Conclusion}

In summary, we have studied the leading order 
effects of scalar and vector diquark
and/or R parity violating couplings on 
hadronic $B$ decays  and $B - {\bar B}$
 mixing.  
 Clearly a diquark must have more than one non-zero couplings to SM fields
 to be able to mediate such processes. We take the economic standpoint 
 that only {\em any two} of such couplings are non-zero. Analysing 
 the present data on $B$-decays, we derive constraints on such pairs that are 
 significantly stronger than those derived from other low energy processes.
Theoretical improvements on nonfactorisation effects and
 estimates of annihilation form factors as well as  precise measurements
of the decay modes at the upcoming
$B$ factories in near future 
 will improve our bounds  on the parameter space for 
diquarks and/or R parity violating couplings in the minimal supersymmetric 
 SM.

\vskip 20pt
\begin{center}
\newpage
{\bf Acknowledgement }
\end{center}
D. Choudhury acknowledges the Department of Science and Technology, India 
for the Swarnajayanti Fellowship grant. 

\vspace*{.2 in}

\newcommand{\ib}[3]   {{\em ibid.\/} {\bf #1} (#3) #2}		       %
\newcommand{\app}[3]  {{\em Acta Phys. Polon.	B\/}{\bf #1} (#3) #2}  %
\newcommand{\ajp}[3]  {{\em Am. J. Phys.\/} {\bf #1} (#3) #2}	       %
\newcommand{\ap}[3]   {{\em Ann. Phys.	(NY)\/}	{\bf #1} (#3)	#2}    %
\newcommand{\araa}[3] {{\em Annu. Rev. Astron. Astrophys.\/}	       %
          {\bf#1} (#3) #2}					       %
\newcommand{\apj}[3]  {{\em Astrophys. J.\/} {\bf #1} (#3) #2}         %
\newcommand{\apjs}[3] {{\em Astrophys. J. Suppl.\/}                    %
          {\bf	#1} (#3) #2}			                       %
\newcommand{\apjl}[3] {{\em Astrophys. J. Lett.\/} {\bf #1} (#3) #2}   %
\newcommand{\astropp}[3]{Astropart. Phys. {\bf #1} (#3) #2}	       %
\newcommand{\eur}[3]  {Eur. Phys. J. {\bf C#1} (#3) #2}                %
\newcommand{\iauc}[4] {{\em IAU Circular\/} #1                         %
       (\ifcase#2\or January \or February \or March  \or April \or May %
                 \or June    \or July     \or August \or September     %
                 \or October \or November \or December                 %
        \fi \ #3, #4)}					               %
\newcommand{\ijmp}[3] {Int. J. Mod. Phys. {\bf A#1} (#3) #2}           %
\newcommand{\jetp}[6] {{\em Zh. Eksp. Teor. Fiz.\/} {\bf #1} (#3) #2   %
     [English translation: {\it Sov. Phys.--JETP } {\bf #4} (#6) #5]}  %
\newcommand{\jetpl}[6]{{\em ZhETF Pis'ma\/} {\bf #1} (#3) #2           %
     [English translation: {\it JETP Lett.\/} {\bf #4} (#6) #5]}       %
\newcommand{\jhep}[3] {JHEP {\bf #1} (#3) #2}                          %
\newcommand{\mpla}[3] {Mod. Phys. Lett. {\bf A#1} (#3) #2}             %
\newcommand{\nat}[3]  {Nature (London) {\bf #1} (#3) #2}	       %
\newcommand{\nuovocim}[3]{Nuovo Cim. {\bf #1} (#3) #2}	               %
\newcommand{\np}[3]   {Nucl. Phys. {\bf B#1} (#3) #2}		       %
\newcommand{\npbps}[3]{Nucl. Phys. B (Proc. Suppl.)                    %
           {\bf #1} (#3) #2}	                                       %
\newcommand{\philt}[3] {Phil. Trans. Roy. Soc. London A {\bf #1} #2    %
	(#3)}							       %
\newcommand{\prev}[3] {Phys. Rev. {\bf #1} (#3) #2}	       	       %
\newcommand{\prd}[3]  {{Phys. Rev.}{\bf D#1} (#3) #2}		       %
\newcommand{\prl}[3]  {Phys. Rev. Lett. {\bf #1} (#3) #2}	       %
\newcommand{\plb}[3]  {{Phys. Lett.} {\bf B#1} (#3) #2}		       %
\newcommand{\prep}[3] {Phys. Rep. {\bf #1} (#3) #2}		       %
\newcommand{\ptp}[3]  {Prog. Theoret. Phys. (Kyoto) {\bf #1} (#3) #2}  %
\newcommand{\rpp}[3]  {Rep. Prog. Phys. {\bf #1} (#3) #2}              %
\newcommand{\rmp}[3]  {Rev. Mod. Phys. {\bf #1} (#3) #2}               %
\newcommand{\sci}[3]  {Science {\bf #1} (#3) #2}		       %
\newcommand{\zp}[3]   {Z.~Phys. C{\bf#1} (#3) #2}		       %
\newcommand{\uspekhi}[6]{{\em Usp. Fiz. Nauk.\/} {\bf #1} (#3) #2      %
     [English translation: {\it Sov. Phys. Usp.\/} {\bf #4} (#6) #5]}  %
\newcommand{\yadfiz}[4]{Yad. Fiz. {\bf #1} (#3) #2 [English	       %
	transl.: Sov. J. Nucl.	Phys. {\bf #1} #3 (#4)]}	       %
\newcommand{\hepph}[1] {(electronic archive:	hep--ph/#1)}	       %
\newcommand{\hepex}[1] {(electronic archive:	hep--ex/#1)}	       %
\newcommand{\astro}[1] {(electronic archive:	astro--ph/#1)}	       %

\newpage


\begin{table}[h]
\begin{center}
\begin{tabular}{|c|c|c|}
\hline
& & \\[-2ex]
Product of Couplings &  Mode &  Allowed Region\\
& & \\[-2ex]
\hline
& & \\[-2ex]
$ h^{(1)}_{13}h_{11}^{(1)}$  & ${\bar B}^0 \to \pi^+ \pi^-$, ${\bar B}^0 \to \pi^0 \pi^0$  &
 $[-2.6 \times 10^{-3}, 1.1 \times 10^{-3}]$\\
& & \\[-2ex]
\hline
& & \\[-2ex]
$ h^{(1)}_{13}h_{12}^{(1)}$  &  ${\bar B}^0 \to D^+ \rho^-$, $B^- \to \pi^-
 {\bar K}^{0*}$, & \\
 & $B^- \to D^0 \rho^-$, ${\bar B} \to {\bar K}^0 \rho^0$ & 
 $[-4.7 \times 10^{-4}, 5.7 \times 10^{-4}]$\\
& & \\[-2ex]
\hline
& & \\[-2ex]
$ h^{(1)}_{23}h_{22}^{(1)}$  & $ {\bar B}^0 \to D^+ D_s^-$, 
 $ {\bar B}^0 \to D^{+*} D_s^-$ &
 $[-5.5 \times 10^{-2}, -5.3 \times 10^{-2}],$\\
& &   $[-6.3 \times 10^{-3},2.2 \times 10^{-3}]$\\
& & \\[-2ex]
\hline
& & \\[-2ex]
$ h^{(1)}_{23}h_{12}^{(1)}$  &  
$ {\bar B}^0 \to K^0 {\bar K}^0$ &
$[-1.4 \times 10^{-3}, 9.5 \times 10^{-4}]$\\
& & \\[-2ex]
\hline
& & \\[-2ex]
$ h^{(1)}_{33}h_{11}^{(1)}$  &  
$ B^0_d - {\bar B}^0_d$ Mixing &
$[-1.4 \times 10^{-7}, -1.3 \times 10^{-7}]$,\\
 & & $[-1.3 \times 10^{-8}, 2.4 \times 10^{-9}]$\\
& & \\
\hline
\end{tabular}
\end{center}
\caption{\em Bounds on $\Phi_1$ couplings in units of ($m_{\Phi_1}/100 \gev$)
 at $90\%$ C.L.}
\label{tab:p1}
\end{table}
\begin{table}[h]
\begin{center}
\begin{tabular}{|c|c|c|}
\hline
& & \\[-2ex]
Product of Couplings &  Mode &  Allowed Region\\
& & \\[-2ex]
\hline
& & \\[-2ex]
$ h^{(2)}_{13}h_{12}^{(2)}$  &  $B^- \to K^- \pi^0$, $B^- \to \pi^-
 {\bar K}^{0}$, & \\
 & $B^- \to K^- \rho^0$ & 
 $[-1.2 \times 10^{-3}, 7.7 \times 10^{-4}]$\\
& & \\[-2ex]
\hline
& & \\[-2ex]
$ h^{(2)}_{23}h_{12}^{(2)}$  &  
$ {\bar B}^0 \to K^0 {\bar K}^0$ &
$[-2.8 \times 10^{-3}, 1.9 \times 10^{-3}]$\\
\hline
\end{tabular}
\end{center}
\caption{\em Bounds on $\Phi_2$ couplings in units of ($m_{\Phi_2}/100 \gev$)
 at $90\%$ C.L.}
\label{tab:p2}
\end{table}

\begin{table}[h]
\begin{center}
\begin{tabular}{|c|c|c|}
\hline
& & \\[-2ex]
Product of Couplings &  Mode &  Allowed Region\\
& & \\[-2ex]
\hline
& & \\[-2ex]
$ h^{(3)}_{13}h_{12}^{(3)}$  & $B^- \to \pi^0 K^-$, 
 $B^- \to K^- \rho^0$, & \\
& ${\bar B}^0 \to D^+ \rho^-$ & 
 $[-1.9 \times 10^{-4}, 2.9 \times 10^{-4}]$\\
& & \\[-2ex]
\hline
& & \\[-2ex]
$ h^{(3)}_{23}h_{12}^{(3)}$  &  
$ B^- \to \pi^0 D_s^-$, ${\bar B}^0 \to \rho^0 J/\Psi$  &
$[-1.5 \times 10^{-3}, 1.7 \times 10^{-3}]$\\
& & \\[-2ex]
\hline
& & \\[-2ex]
$ {\tilde h}^{(3)}_{13}{\tilde h}_{11}^{(3)}$ 
 & ${\bar B}^0 \to \pi^0 \pi^0$, $B^- \to \pi^- \pi^0$  &
 $[-3.3 \times 10^{-3}, 2.9 \times 10^{-3}]$\\
& & \\[-2ex]
\hline
& & \\[-2ex]
$ {\tilde h}^{(3)}_{13}{\tilde h}_{12}^{(3)}$  & ${\bar B}^0 \to \pi^0 
{\bar K}^{0*}$, 
$ B^- \to \pi^0 K^-$ & $[-1.1 \times 10^{-3}, 7.8 \times 10^{-4}]$\\
& & \\[-2ex]
\hline
& & \\[-2ex]
$ {\tilde h}^{(3)}_{13}{\tilde h}_{21}^{(3)}$  & $ B^- \to \pi^- D^0$, 
$ B^- \to \rho^- D^0$, &  \\
 &  ${\bar B}^0 \to D^0 \pi^0 $  &
 $[-1.2 \times 10^{-3}, 1.4 \times 10^{-3}]$\\
& & \\[-2ex]
\hline
& & \\[-2ex]
$ {\tilde h}^{(3)}_{23}{\tilde h}_{22}^{(3)}$  & $ B^- \to D^0 D_s^{-*}$, 
$ {\bar B}^0 \to D^+ D_s^-$ 
&  $[-8.7 \times 10^{-3}, 3.1 \times 10^{-2}]$\\
& & \\[-2ex]
\hline
& & \\[-2ex]
$ {\tilde h}^{(3)}_{23}{\tilde h}_{21}^{(3)}$  &  
$ {\bar B}^0 \to \pi^0 J/\Psi$, $  {\bar B}^0 \to \rho^0 J/\Psi$ &
$[-4.2 \times 10^{-3}, 4.2 \times 10^{-3}]$\\
& & \\[-2ex]
\hline
& & \\[-2ex]
$ {\tilde h}^{(3)}_{23}{\tilde h}_{12}^{(3)}$  &  
$ B^- \to D_s^- \pi^0$ &
$[-1.4 \times 10^{-2}, 2.0 \times 10^{-2}]$\\
& & \\[-2ex]
\hline
& & \\[-2ex]
$ h^{(3)}_{13}{\tilde h}^{(3)}_{11}$  & ${\bar B}^0 \to \pi^0 \pi^0$, ${\bar B}^0 \to 
 \pi^+ \pi^-$  & $[-5.1 \times 10^{-3}, 7.4 \times 10^{-3}]$\\
& & \\[-2ex]
\hline
& & \\[-2ex]
$ h^{(3)}_{12}{\tilde h}^{(3)}_{13}$ & $B^- \to K^- \pi^0$, ${\bar B}^0
 \to \pi^+ K^-$, & \\
& $B^- \to D^0 \pi^-$ & $[-1.4 \times 10^{-3},
 2.2 \times 10^{-3}]$\\
& & \\[-2ex]
\hline
& & \\[-2ex]
$ h^{(3)}_{13}{\tilde h}^{(3)}_{12}$ & $B^- \to K^- \pi^0$, ${\bar B}^0
 \to \pi^+ K^-$, & $[-3.3 \times 10^{-3},
 1.4 \times 10^{-3}]$ ,\\
& & \\[-2ex]
\hline
& & \\[-2ex]
$ h^{(3)}_{13}{\tilde h}^{(3)}_{21}$ & ${\bar B}^0 \to D^0 \pi^0$,
 $B^- \to D^0 \pi^-$,  & \\
& $B^- \to D^0 \rho^-$ & $[-5.2 \times 10^{-3},
 2.2 \times 10^{-3}]$\\
& & \\[-2ex]
\hline
& & \\[-2ex]
$ h^{(3)}_{23}{\tilde h}^{(3)}_{22}$ & ${\bar B}^0 \to D^+ D^-_s$, $B^- \to D^0 D^-_s$, & \\
&  ${\bar B}^0 \to D^+ D_s^{-*}$  & $[-4.9 \times 10^{-3},
 1.5 \times 10^{-2}]$\\
& & \\[-2ex]
\hline
& & \\[-2ex] 
$ h^{(3)}_{12}{\tilde h}^{(3)}_{23}$ & $B^- \to \pi^0 D^-_s$ 
& $[-1.6 \times 10^{-2},
 1.1 \times 10^{-2}]$\\
& & \\[-2ex]
\hline
& & \\[-2ex] 
$ h^{(3)}_{23}{\tilde h}^{(3)}_{12}$ & $B^- \to \pi^0 D^-_s$ 
& $[-1.1 \times 10^{-2},
 1.6 \times 10^{-2}]$\\
& & \\
\hline
\end{tabular}
\end{center}
\caption{\em Bounds on $\Phi_3$ couplings in units of ($m_{\Phi_3}/100 \gev$)
 at $90\%$ C.L.}
\label{tab:p3}
\end{table}
\begin{table}[h]
\begin{center}
\begin{tabular}{|c|c|c|}
\hline
& & \\[-2ex]
Product of Couplings &  Mode &  Allowed Region\\
& & \\[-2ex]
\hline
& & \\[-2ex]
$ h^{(4)}_{13}h_{11}^{(4)}$  & ${\bar B}^0 \to \pi^+ \pi^-$, ${\bar B}^0 \to \pi^0 \pi^0$  &
 $[-1.1 \times 10^{-3}, 2.6 \times 10^{-3}]$\\
& & \\[-2ex]
\hline
& & \\[-2ex]
$ h^{(4)}_{13}h_{12}^{(4)}$  & ${\bar B}^0 \to \pi^+ K^-$, 
 ${\bar B}^0_s \to K^+ K^-$, &\\
& ${\bar B}^0 \to {\bar K}^0 \rho^0$ & 
 $[-8.1 \times 10^{-4}, 5.4 \times 10^{-4}]$\\
& & \\[-2ex]
\hline
& & \\[-2ex]
$ h^{(4)}_{23}h_{22}^{(4)}$  & $ {\bar B}^0 \to D^+ D_s^-$, 
$ {\bar B}^0 \to D^{+*} D_s^-$ & 
 $[-2.2 \times 10^{-3}, 6.5 \times 10^{-3}],$\\
& $B^- \to D^0 D^-_s$ & 
 $[5.3 \times 10^{-2}, 5.5  \times 10^{-2}]$\\
& & \\[-2ex]
\hline
& & \\[-2ex]
$ h^{(4)}_{23}h_{12}^{(4)}$  &  
$ B^- \to \pi^0 D_s^-$,$ {\bar B}^0 \to \rho^0 J/\Psi$ &
$[-3.4 \times 10^{-3}, 3.1 \times 10^{-3}]$\\
& & \\[-2ex]
\hline
& & \\[-2ex]
$ {\tilde h}^{(4)}_{13}{\tilde h}_{11}^{(4)}$ 
 & ${\bar B}^0 \to \pi^0 \pi^0$, $B^- \to \pi^- \pi^0$  &
 $[-3.3 \times 10^{-3}, 2.9 \times 10^{-3}]$\\
& & \\[-2ex]
\hline
& & \\[-2ex]
$ {\tilde h}^{(4)}_{13}{\tilde h}_{12}^{(4)}$  & ${\bar B}^0 \to \pi^0 
{\bar K}^{0*}$, 
$ B^- \to \pi^0 K^-$ & $[-1.1 \times 10^{-3}, 7.8 \times 10^{-4}]$\\
& & \\[-2ex]
\hline
& & \\[-2ex]
$ {\tilde h}^{(4)}_{13}{\tilde h}_{21}^{(4)}$  & $ B^- \to \pi^- D^0$, 
$ B^- \to \rho^- D^0$, &  \\
 &  ${\bar B}^0 \to D^0 \pi^0 $  &
 $[-1.2 \times 10^{-3}, 1.4 \times 10^{-3}]$\\
& & \\[-2ex]
\hline
& & \\[-2ex]
$ {\tilde h}^{(4)}_{23}{\tilde h}_{22}^{(4)}$  & $ B^- \to D^0 D_s^{-*}$, 
$ {\bar B}^0 \to D^+ D_s^-$ 
&  $[-8.7 \times 10^{-3}, 3.1 \times 10^{-2}]$\\
& & \\[-2ex]
\hline
& & \\[-2ex]
$ {\tilde h}^{(4)}_{23}{\tilde h}_{21}^{(4)}$  &  
$ {\bar B}^0 \to \pi^0 J/\Psi$, $  {\bar B}^0 \to \rho^0 J/\Psi$ &
$[-4.2 \times 10^{-3}, 4.2 \times 10^{-3}]$\\
& & \\[-2ex]
\hline
& & \\[-2ex]
$ {\tilde h}^{(4)}_{23}{\tilde h}_{12}^{(4)}$  &  
$ B^- \to D_s^- \pi^0$ &
$[-1.4 \times 10^{-2}, 2.0 \times 10^{-2}]$\\
& & \\[-2ex]
\hline
& & \\[-2ex]
$ h^{(4)}_{11}{\tilde h}^{(4)}_{13}$ & ${\bar B}^0 \to \pi^+ \pi^-$, 
${\bar B}^0 \to \pi^0 \pi^0$ 
 & $[-3.6 \times 10^{-3}, 6.4 \times 10^{-3}]$\\
& & \\[-2ex]
\hline
& & \\[-2ex]
$ h^{(4)}_{13}{\tilde h}^{(4)}_{11}$ & ${\bar B}^0 \to \pi^+ \pi^-$, 
${\bar B}^0 \to \pi^0 \pi^0$ 
 & $[-6.4 \times 10^{-3}, 3.6 \times 10^{-3}]$\\
& & \\[-2ex]
\hline
& & \\[-2ex]
$ h^{(4)}_{12}{\tilde h}^{(4)}_{13}$ & ${\bar B}^0_s \to K^+ K^-$, ${\bar B}^0
 \to \pi^+ K^-$ & $[-1.5 \times 10^{-2}, -1.2 \times 10^{-2}$\\
& & $[-2.3 \times 10^{-3},
 1.5 \times 10^{-3}]$\\
& & \\[-2ex]
\hline
& & \\[-2ex]
$ h^{(4)}_{13}{\tilde h}^{(4)}_{12}$ & ${\bar B}^0_s \to K^+ K^-$,
 ${\bar B}^0
 \to \pi^+ K^-$, & \\
& ${\bar B}^0 \to {\bar K}^{0*} \pi^0$, $B^- \to K^- \rho^0$  & 
$[-1.5 \times 10^{-3}, 2.3 \times 10^{-3}\}]$\\
& & \\[-2ex]
\hline
& & \\[-2ex]
 $h^{(4)}_{13}{\tilde h}^{(4)}_{21}$ & ${\bar B}^0 \to D^0 \pi^0$, $ B^-
 \to D^0 \rho^-$  & $[-3.7 \times 10^{-3}, 2.2 \times 10^{-2}]$\\
& & \\[-2ex]
\hline
& & \\[-2ex]
$ h^{(4)}_{22}{\tilde h}^{(4)}_{23}$ & ${\bar B}^0 \to D^+ D^-_s$, $B^- \to D^0 D^-_s$, & $[-3.5 \times 10^{-3}, 1.0 \times 10^{-2}]$\\
&  & $[7.4 \times 10^{-2}, 8.8 \times 10^{-2}]$\\
& & \\[-2ex]
\hline
& & \\[-2ex] 
$ h^{(4)}_{23}{\tilde h}^{(4)}_{22}$ & ${\bar B}^0 \to D^+ D^-_s$, $B^- \to D^0 D^-_s$, & \\
&  ${\bar B}^0 \to D^{+*} D_s^{-}$  & $[-1.0 \times 10^{-2}, 3.5 \times 10^{-3}]$\\
& & \\[-2ex]
\hline
& & \\[-2ex] 
$ h^{(4)}_{12}{\tilde h}^{(4)}_{23}$ & $B^- \to \pi^0 D^-_s$ 
& $[-8.0 \times 10^{-3}, 1.2 \times 10^{-2}]$\\
& & \\[-2ex]
\hline
& & \\[-2ex] 
$ h^{(4)}_{23}{\tilde h}^{(4)}_{12}$ & $B^- \to \pi^0 D^-_s$ 
& $[-1.2 \times 10^{-2}, 8.0 \times 10^{-3}]$\\
& & \\
\hline
\end{tabular}
\end{center}
\caption{\em Bounds on $\Phi_4$ couplings in units of ($m_{\Phi_4}/100 \gev$)
 at $90\%$ C.L.}
\label{tab:p4}
\end{table}
\begin{table}[h]
\begin{center}
\begin{tabular}{|c|c|c|}
\hline
Product of Couplings &  Mode &  Allowed Region\\
\hline
& & \\[-2ex]
$ h^{(7)}_{13}h_{11}^{(7)}$  & ${\bar B}^0 \to \pi^0 \pi^0$  &
 $[-1.1 \times 10^{-3}, 1.3 \times 10^{-3}]$\\
& & \\[-2ex]
\hline
& & \\[-2ex]
$ h^{(7)}_{13}h_{12}^{(7)}$  & 
$B^- \to \pi^- {\bar K}^0$, 
$ B^- \to \pi^0 K^-$ & \\
& $  {\bar B}^0 \to \pi^0 {\bar K}^0$ &
  $[-3.9 \times 10^{-4}, 7.3 \times 10^{-4}]$\\
& & \\[-2ex]
\hline
& & \\[-2ex]
$ h^{(7)}_{23}h_{12}^{(7)}$ &  
$  {\bar B}^0 \to K^0 {\bar K}^0$ &
$[-9.5 \times 10^{-4}, 1.4 \times 10^{-3}]$\\
& & \\[-2ex]
\hline
& & \\[-2ex]
$ h^{(7)}_{33}h_{11}^{(7)}$ &  
$B^0_d - {\bar B}^0_d$ Mixing &
$[-1.4 \times 10^{-7}, -1.3 \times 10^{-7}]$,\\
 & & $[-1.3 \times 10^{-8}, 2.4 \times 10^{-9}]$\\
& & \\
\hline
\end{tabular}
\end{center}
\caption{\em Bounds on $\Phi_7$ couplings in units of ($m_{\Phi_7}/100 \gev$)
 at $90\%$ C.L.}
\label{tab:p7}
\end{table}
\begin{table}[h]
\begin{center}
\begin{tabular}{|c|c|c|}
\hline
Product of Couplings &  Mode &  Allowed Region\\
\hline
& & \\[-2ex]
$ h^{(8)}_{13}h_{12}^{(8)}$  & 
$B^- \to \pi^- {\bar K}^0$, 
$ B^- \to \pi^0 K^-$ & \\
& $  B^- \to \pi^- {\bar K}^{0*}$ &
  $[-7.9 \times 10^{-4}, 1.2 \times 10^{-3}]$\\
& & \\[-2ex]
\hline
& & \\[-2ex]
$ h^{(8)}_{23}h_{12}^{(8)}$ &  
$  {\bar B}^0 \to K^0 {\bar K}^0$ &
$[-1.9 \times 10^{-3}, 2.8 \times 10^{-3}]$\\
& & \\
\hline
\end{tabular}
\end{center}
\caption{\em Bounds on $\Phi_8$ couplings in units of ($m_{\Phi_8}/100 \gev$)
 at $90\%$ C.L.}
\label{tab:p8}
\end{table}
\begin{table}[htbp]
\begin{center}
\begin{tabular}{|c|c|c|}
\hline
& & \\[-2ex]
Product of Couplings &  Mode &  Allowed Region\\
& & \\[-2ex]
\hline
& & \\[-2ex]
$ \vartheta^{(1)}_{13}\vartheta_{11}^{(1)}$  & ${\bar B}^0 \to \pi^+ \pi^-$,
 ${\bar B}^0 \to \rho^0 \rho^0$
 & $[-6.2\times 10^{-3}, 3.2\times 10^{-3}]$\\
& & \\[-2ex]
\hline
& & \\[-2ex]
$ \vartheta^{(1)}_{31}\vartheta_{11}^{(1)}$ & ${\bar B}^0 \to \pi^0 \pi^0$  &
 $[-2.1\times 10^{-3}, 1.7 \times 10^{-3}]$\\
& & \\[-2ex]
\hline
& & \\[-2ex]
$ \vartheta^{(1)}_{13}\vartheta_{12}^{(1)}$ & ${\bar B}^0 \to \pi^+ K^-$,
 $B^- \to K^- \pi^0$, & 
 $[-1.4 \times 10^{-3}, 2.0 \times 10^{-3}],$\\
& ${\bar B}^0 \to \rho^+ K^-$ &
 $ [1.0 \times 10^{-2}, 1.1 \times 10^{-2}]$\\
& & \\[-2ex]
 \hline
& & \\[-2ex]
$ \vartheta^{(1)}_{13}\vartheta_{21}^{(1)}$ & $B^- \to \pi^- {\bar K}^0$,
$B^- \to \pi^- {\bar K}^{0*}$  &
 $[-1.0 \times 10^{-3}, -2.6 \times 10^{-3}]$,\\
& $B^- \to D^0 \rho^-$ &  $[-8.1 \times 10^{-4}, 9.8 \times 10^{-4}]$\\
& & \\[-2ex]
\hline
& & \\[-2ex]
$ \vartheta^{(1)}_{31}\vartheta_{12}^{(1)}$ & ${\bar B}^0 \to {\bar K}^0 \pi^0$,
$B^- \to \pi^- {\bar K}^0$ & \\
 & $B^- \to \pi^- {\bar K}^{0*}$  &
 $[-9.8 \times 10^{-4}, 8.1 \times 10^{-4}]$\\
& & \\[-2ex]
 \hline
& & \\[-2ex]
$ \vartheta^{(1)}_{31}\vartheta_{21}^{(1)}$ & 
$B^- \to \pi^0 K^-$,
  $B^- \to \rho^0 K^-$ &
 $[-1.5 \times 10^{-3}, 5.3 \times 10^{-4}]$\\
& & \\[-2ex]
  \hline
& & \\[-2ex]
$ \vartheta^{(1)}_{23}\vartheta_{22}^{(1)}$ & ${\bar B}^0 \to D^+ D_s^-$,
 $B^- \to D^0 D_s^-$ &\\
 &  $B^- \to D^{0*} D_s^-$
  & $[-9.2 \times 10^{-3}, 3.1 \times 10^{-3}]$\\
& & \\[-2ex]
\hline
& & \\[-2ex]
$ \vartheta^{(1)}_{23}\vartheta_{12}^{(1)}$ & ${\bar B}^0 \to K^0 {\bar K}^0$
  & $[-1.3 \times 10^{-3}, 1.9 \times 10^{-3}]$\\
& & \\[-2ex]
\hline
& & \\[-2ex]
$ \vartheta^{(1)}_{23}\vartheta_{21}^{(1)}$ & $B^0 \to \pi^0 J/\Psi$,
$B^0 \to \rho^0 J/\Psi$, &
 $[-2.1 \times 10^{-3}, 2.1 \times 10^{-3}]$\\
& & \\[-2ex]
\hline
& & \\[-2ex]
$ \vartheta^{(1)}_{32}\vartheta_{12}^{(1)}$ & $B^0 \to K^0 {\bar K}^0$ &
 $[-4.7 \times 10^{-3}, 7.0 \times 10^{-3}]$\\
& & \\[-2ex]
 \hline
& & \\[-2ex]
$ \vartheta^{(1)}_{32}\vartheta_{21}^{(1)}$ & $B^0 \to K^0 {\bar K}^0$ &
 $[-1.9 \times 10^{-3}, 1.3 \times 10^{-3}]$\\
 & & \\
[-2ex]
 \hline
& & \\[-2ex]
$ \vartheta^{(1)}_{33}\vartheta_{11}^{(1)}$ & $B^0_d -{\bar B}^0_d$ Mixing  &
 $[-5.4 \times 10^{-8}, -4.8 \times 10^{-8}]$,\\
  & & $[-4.9 \times 10^{-9}, 8.9 \times 10^{-10}]$\\
 & &\\
\hline
\end{tabular}
\end{center}
\caption{\em Bounds on $V_1$ couplings in units of ($m_{V_1}/100 \gev$)
 at $90\%$ C.L.}
\label{tab:v1}
\end{table}
\begin{table}[htbp]
\begin{center}
\begin{tabular}{|c|c|c|}
\hline
Product of Couplings &  Mode &  Allowed Region\\
& & \\[-2ex]
\hline
& & \\[-2ex]
$ \vartheta^{(3)}_{13}\vartheta_{11}^{(3)}$  & ${\bar B}^0 \to \pi^+ \pi^-$
 & $[-3.2 \times 10^{-3}, 1.0 \times 10^{-2}]$\\
& & \\[-2ex]
\hline
& & \\[-2ex]
$ \vartheta^{(3)}_{31}\vartheta_{11}^{(3)}$ & ${\bar B}^0 \to \pi^0 \pi^0$  &
 $[-1.4 \times 10^{-3}, 1.8 \times 10^{-3}]$\\
& & \\[-2ex]
\hline
& & \\[-2ex]
$ \vartheta^{(3)}_{13}\vartheta_{12}^{(3)}$ & ${\bar B}^0 \to \pi^+ K^-$,
 ${\bar B}^0 \to \rho^+ K^-$ &
 $[-1.2 \times 10^{-2}, -1.0 \times 10^{-2}],$\\
& & $ [-2.0 \times 10^{-3}, 1.4 \times 10^{-3}]$\\
& & \\[-2ex]
 \hline
& & \\[-2ex]
$ \vartheta^{(3)}_{13}\vartheta_{21}^{(3)}$ & $B^- \to D^0 \rho^-$, 
${\bar B}^0 \to D^{0*} \rho^0$ &
 $[-1.0 \times 10^{-3}, 9.0 \times 10^{-3}]$\\
& & \\[-2ex]
\hline
& & \\[-2ex]
$ \vartheta^{(3)}_{31}\vartheta_{12}^{(3)}$ & ${\bar B}^0 \to D^0 \pi^0$,
 $B^- \to D^0 \pi^-$  &
 $[-1.5 \times 10^{-3}, 4.1 \times 10^{-3}]$\\
& & \\[-2ex]
 \hline
& & \\[-2ex]
$ \vartheta^{(3)}_{31}\vartheta_{21}^{(3)}$ & 
$B^- \to \pi^0  K^-$, $B^- \to K^- \rho^0$  &
 $[-7.3 \times 10^{-4}, 1.7 \times 10^{-3}]$\\
& & \\[-2ex]
 \hline
& & \\[-2ex]
$ \vartheta^{(3)}_{32}\vartheta_{22}^{(3)}$ & ${\bar B}^0 \to D^{+} D_s^-$,
$B^- \to D^{0} D_s^-$ & $[-1.5 \times 10^{-3}, 4.3 \times 10^{-3}]$\\
& & \\[-2ex]
\hline
& & \\[-2ex]
$ \vartheta^{(3)}_{23}\vartheta_{22}^{(3)}$ & 
$B^- \to D^{0*} D_s^-$ & $[-2.2 \times 10^{-1}, -1.4 \times 10^{-1}]$,\\
&  &
$ [-2.0 \times 10^{-2}, 5.7 \times 10^{-2}]$\\
& & \\[-2ex]
\hline
& & \\[-2ex]
$ \vartheta^{(3)}_{23}\vartheta_{12}^{(3)}$ & ${\bar B}^0 \to D_s^- \pi^+$
  & $[-1.1 \times 10^{-2}, 1.4 \times 10^{-2}]$\\
& & \\[-2ex]
\hline 
& & \\[-2ex]
$ \vartheta^{(3)}_{23}\vartheta_{21}^{(3)}$ & ${ \bar B}^0 \to \rho^0 J/\Psi$ &
 $[-2.5 \times 10^{-3}, 2.1 \times 10^{-3}]$\\
& & \\[-2ex]
\hline
& & \\[-2ex]
$ \vartheta^{(3)}_{32}\vartheta_{12}^{(3)}$ & ${ \bar B}^0 \to \pi^0 J/\Psi$ &
 $[-2.5 \times 10^{-3}, 2.1 \times 10^{-3}]$\\
& & \\[-2ex]
\hline
& & \\[-2ex]
$ \vartheta^{(3)}_{32}\vartheta_{21}^{(3)}$ & $B^- \to D_s^- \pi^0 $ &
 $[-7.0 \times 10^{-3}, 1.0 \times 10^{-2}]$\\
& & \\
 \hline
\end{tabular}
\end{center}
\caption{\em Bounds on $V_3$ couplings in units of ($m_{V_3}/100 \gev$)
 at $90\%$ C.L.}
\label{tab:v3}
\end{table}

\end{document}